\documentclass[a4paper]{article}

\usepackage{INTERSPEECH2021}

\usepackage{siunitx}
\usepackage{multicol,multirow}
\usepackage{url}
\usepackage{comment}
\usepackage{capt-of} 
\usepackage[subrefformat=parens]{subcaption}
\usepackage{booktabs}
\title{Investigation for Relative Voice Impression Estimation}
\name{Kenichi Fujita and Yusuke Ijima}
\address{
  NTT, Inc., Japan}
\email{kenichi.fujita@ntt.com}

\begin{document}
\setlength\textfloatsep{7pt} 
\setlength\dbltextfloatsep{7pt} 
\setlength\floatsep{7pt} 
\setlength\abovecaptionskip{2pt} 
\setlength\belowcaptionskip{2pt} 
\captionsetup[subfloat]{aboveskip=2pt,belowskip=4pt} 

\maketitle
\begin{abstract}
Paralinguistic and non-linguistic aspects of speech strongly influence listener impressions. 
While most research focuses on absolute impression scoring, this study investigates relative voice impression estimation (RIE), a framework for predicting the perceptual difference between two utterances from the same speaker.
The estimation target is a low-dimensional vector derived from subjective evaluations, quantifying the perceptual shift of the second utterance relative to the first along an antonymic axis (e.g., ``Dark--Bright'').
To isolate expressive and prosodic variation, we used recordings of a professional speaker reading a text in various styles.
We compare three modeling approaches: classical acoustic features commonly used for speech emotion recognition, self-supervised speech representations, and multimodal large language models~(MLLMs).
Our results demonstrate that models using self-supervised representations outperform methods with classical acoustic features, particularly in capturing complex and dynamic impressions (e.g., ``Cold--Warm'') where classical features fail.
In contrast, current MLLMs prove unreliable for this fine-grained pairwise task.
This study  provides the first systematic investigation of RIE and demonstrates the strength of self-supervised speech models in capturing subtle perceptual variations.
\end{abstract}

\section{Introduction}

Human speech conveys far more than lexical content. 
For instance, paralinguistic and non-linguistic information, such as speech rate, breathiness, and prosodic patterns, strongly shape listeners' impressions~\cite{cowen2019mapping,scherer1986vocal,murray1993toward}. 
Decades of research have explored relationships between acoustic features and perceptual attributes such as emotion~\cite{schuller2013computational}, speaker traits~\cite{6834774, app11188776}, and perceived audio quality~\cite{pandey25_ssw}.
This line of work has led to various automatic estimation tasks, including speech emotion recognition (SER)~\cite{wani21_Access, 10711189}, speaker-impression assessment~\cite{fujita25_interspeech}, and prediction of subjective quality ratings~\cite{Erica_Cooper2024e24.12}.
However, most existing studies adopt an \emph{absolute} formulation that maps a single audio input to a scalar label or score.
In many practical scenarios, however, \emph{relative} perceptual differences are equally or more important.

People naturally perceive and control voice impressions in relative terms.
In everyday interaction, we compare voices, for example, noticing when someone sounds more energetic than yesterday or slightly less confident than usual.
A similar pattern occurs in creative contexts such as voice acting~\cite{kanagawa25_ssw}, where direction is given relative to a reference, for instance ``more anger than the previous take'' or ``slightly softer than before.''
In both cases, the essential information lies in how impressions change relative to a baseline rather than in their absolute levels.

Despite the importance of such relative perceptions, most computational models for speech impressions, emotion, or naturalness still rely on absolute scoring. 
A simple workaround is to compute independent scores for two samples and take their differences, but this approach ignores how humans perceive change directly.
Subsequent attempts, such as the Voice
Timbre Attribute Detection (VTaD) 2025 Challenge~\cite{sheng2025voicetimbreattributedetection}, have begun to explore pairwise comparisons by asking which of two utterances exhibits a stronger attribute. 
However, this formulation is limited to binary judgment and does not capture the magnitude of perceptual differences. 
Furthermore, existing efforts lacked a systematic investigation into how such relative perceptual differences can be modeled from speech.

We address this gap by introducing \emph{relative voice impression estimation (RIE)}.
Given two utterances of reading the same text by the same speaker with \emph{different speaking styles}, our goal is to estimate both the \emph{direction} and \emph{magnitude} of the perceived impression change. 
Formally, let $x_a$ and $x_b$ denote two speech signals from the same speaker reading the same text. 
RIE learns
\begin{equation}
\mathbf{r_{rel}} = f(x_a, x_b) \in \mathbb{R}^K,
\end{equation}
where $\mathbf{r_{rel}}$ represents a low-dimensional vector of perceptual differences in voice impressions derived from subjective evaluations. 
Each dimension quantifies how the impression of $x_b$ is perceived to shift relative to $x_a$ along an antonymic axis, for example ``dark--bright.''
This formulation enables direct modeling of perceptual change from paired utterances, reflecting how listeners perceive subtle shifts in voice impressions.

In this paper, we present the first systematic study of RIE using three complementary modeling approaches. 
We compare (1)~interpretable acoustic features from openSMILE~\cite{eyben_opensmile}, (2)~representations from self-supervised learning (SSL) models that have shown to be effective on speech-related tasks~\cite{9964237,saeki22c_interspeech,9893562}, and (3)~multimodal large language models (MLLMs)~\cite{wu23_mllm,yin24_mllm}.
While the first two focus on acoustic modeling, MLLMs allow us to examine whether language-grounded audio-text models can capture relative perceptual differences between utterances. 
Recent studies have demonstrated the capability of MLLMs in multimodal understanding,~\cite{NEURIPS2024_c7f43ada, murzaku2025omnivoxzeroshotemotionrecognition, jia2025interpretableaudioeditingevaluation}, suggesting their potential to capture paralinguistic cues relevant to voice impressions.
Our contributions are as follows:
(1) we formalize the RIE task, and
(2) we provide the first systematic comparison across acoustic features, SSL representations, and MLLM approaches.

\begin{figure}[tb]
  \centering
  \includegraphics[width=0.9\linewidth]{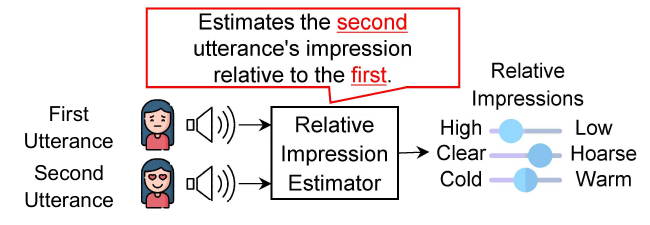}
  \caption{Overview of relative voice impression estimation. The system takes two utterances and outputs a vector of perceptual differences along predefined dimensions.}
  \label{fig:overview_rie}
\end{figure}

\section{Dataset and impression representation}
\subsection{Dataset}\label{sect:dataset}
We used an internal Japanese speech dataset sampled at \SI{22}{\kHz}.
The dataset includes recordings from a single professional female voice actor.
To ensure experimental control, she repeatedly read the same text with different speaking styles.
Using a single speaker eliminates the variability in fundamental frequency range and vocal timbre that is inherent to multi-speaker data.
This setup allows us to isolate and analyze purely expressive differences in both acoustic and latent representations, providing a foundation for future multi-speaker extensions.

The recordings were produced in accordance with the Guideline for TTS Speaking Style Classification~\cite{JEITA}, which defines 55 speaking styles such as ``sleepy,'' ``child-oriented,'' and ``urgent.''
Among these, we used 52 styles, excluding ``neutral,'' ``screaming,'' and ``laughing''.
The dataset comprised 814 utterance pairs derived from 1,087 Japanese utterances, totaling 3,372 seconds of speech.
All recordings were downsampled to \SI{16}{\kHz} for model input, including feature extraction with openSMILE and SSL models.

\subsection{Impression difference vector}
The impression difference vector is a low-dimensional representation.
Each dimension quantifies the degree of perceived contrast along an antonymic impression pair derived from subjective evaluations.
We use the following nine pairs of descriptors used in voice quality assessment~\cite{fujita25_interspeech,Mizuki_Nagano2024e24.14,kasuya1999_extraction}: A) High--Low pitched,  B) Clear--Hoarse, C) Calm--Restless, D) Powerful--Weak, E) Youthful--Elderly, F) Thick--Thin, G) Tense--Relaxed, H) Dark--Bright, and I) Cold--Warm. 
The Female--Male dimension used in previous studies was excluded because all utterances were produced by the same female speaker.
Each dimension represents a relative impression, indicating how the second utterance is perceived relative to the first.

Subjective evaluations were conducted via crowdsourcing.
We recruited 3,920 participants, and each utterance pair was evaluated on the nine impression dimensions using a seven-point Likert scale (e.g., 1 = dark, 7 = bright, 4 = no difference).
Each pair was rated by at least ten participants, and the mean rating per dimension was used to form the final impression difference vector $\mathbf{r_{rel}}$.
To mitigate order effects, all utterance pairs were presented in both AB and BA orders.

\section{Method and experimental conditions}
For methods that can be trained on our dataset (Sections~\ref{sect:classical} and~\ref{sect:ssl_based}), we conducted 10-fold cross-validation to ensure robust evaluation.
In contrast, commercial systems, i.e., MLLMs, were evaluated without retraining, as their adaptation typically requires much larger datasets. 
To maintain consistency, these models were tested using one fold from the same cross-validation setup.
These MLLMs were evaluated in a zero-shot manner without any fine-tuning on our data.
This experiment was exploratory rather than a strict benchmark, and the aim was to examine the potential of recent audio-aware LLMs for perceptual reasoning.

\begin{table}[tb]
 \centering
  \begin{minipage}{0.9\linewidth} 
    \caption{Hyperparameters for regressors. }
    \setlength{\tabcolsep}{2pt}
    \label{table:hyperparameters}
    \centering
    \resizebox{\linewidth}{!}{%
    \begin{tabular}{@{}c@{ }l*{1}{r}@{}}
    \toprule
      \multicolumn{1}{c}{Method} & \multicolumn{1}{c}{Parameters}\\ 
      \midrule
      Ridge& $\alpha$ = $0.5$\\ 
      PLS2& number of components = $5$\\ 
      RF& number of estimators = $300$\\ 
      GBDT& \begin{tabular}{l}number of estimators = $100$, max depth = $3$, \\criterion = Friedman MSE\end{tabular}\\ 
      SVR& C = $10.0$, $\epsilon$ = $0.1$, kernel = RBF, $\gamma$ = scale\\ 
    \bottomrule
    \end{tabular}%
    }
  \end{minipage}
\end{table}

\begin{figure}[tb]
  \centering
  \includegraphics[width=0.7\linewidth]{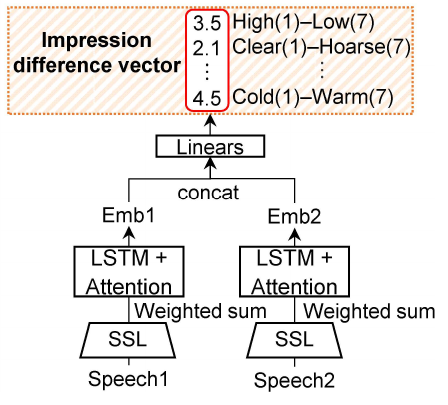}
  \caption{Overview of the SSL-based estimation.}
  \label{fig:overview_propose}
\end{figure}

\begin{table*}[tb]
  \centering
  \begin{minipage}{0.48\linewidth}
    \centering
    \includegraphics[width=\linewidth]{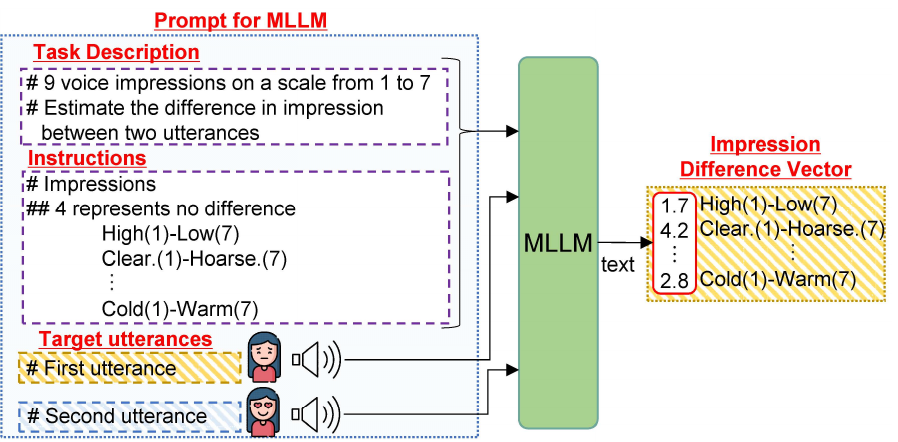}
    \captionof{figure}{Overview of the MLLM-based estimation.}
    \label{fig:overview_mllm}
  \end{minipage}\hfill
  \begin{minipage}{0.51\linewidth}
    \centering
    \captionof{table}{Top-8 feature–target Pearson correlations used for feature selection.
    F0 related features are F0semitoneFrom27.5Hz, MFCC1 and 2 represent mfcc1 and mfcc2,
    $\alpha$, HB, ${F1}_{BW}$, and SF represents alphaRatio, hammarbergIndexV, F1bandwidth and spectralFluxV.
    Bold denotes top-8 features.}
    \label{table:top_features}
    \setlength{\tabcolsep}{2pt}
    \resizebox{\linewidth}{!}{%
      \begin{tabular}{cl*{10}{S}}
        \toprule
      \multicolumn{2}{c}{} &
      \multicolumn{4}{c}{F0} &
      \multicolumn{1}{c}{MFCC1} &
      \multicolumn{1}{c}{MFCC2} &
      \multicolumn{1}{c}{$\alpha$} &
      \multicolumn{1}{c}{HB} &
      \multicolumn{1}{c}{$\mathrm{F1_{BW}}$} &
      \multicolumn{1}{c}{SF} \\
      \cmidrule(r){3-6} \cmidrule(r){7-8} \cmidrule(r){9-9} \cmidrule(r){10-10} \cmidrule(r){11-11} \cmidrule(lr){12-12}
      \multicolumn{2}{c}{} &
      {mean} & {20} & {50} & {80} & {mean} & {mean} &{mean}&{mean}&{mean}&{mean}\\
    \midrule

    A & High--Low         & \textbf{\textminus0.50} & \textbf{\textminus0.42} & \textbf{\textminus0.48} & \textbf{\textminus0.50} &  \textbf{0.50
    }&  \textbf{0.49} & \textbf{\textminus0.47} &  \textbf{0.39} &  0.31 & \textminus0.26 \\
    B & Clear--Hoarse     &   \textbf{0.50} &   \textbf{0.43} &   \textbf{0.49} &   \textbf{0.50} &  \textbf{\textminus0.48} &  \textbf{\textminus0.49} &   \textbf{0.42} &  \textbf{\textminus0.38} & \textminus0.28 &   0.26 \\
    C & Calm--Restless    &   \textbf{0.25} &  0.17 &   \textbf{0.24} &   \textbf{0.28} &  \textbf{\textminus0.31} &  \textbf{\textminus0.25} &   \textbf{0.33} &  \textbf{\textminus0.32} &  \textbf{\textminus0.19} &  0.16 \\
    D & Powerful--Weak    &  \textbf{\textminus0.30} & \textminus0.22 &  \textbf{\textminus0.27} &  \textbf{\textminus0.35} &   \textbf{0.32} &   \textbf{0.37} &  \textbf{\textminus0.35} &   \textbf{0.32} &  0.16 &  \textbf{\textminus0.34} \\
    E & Youthful--Elderly &  \textbf{\textminus0.52} &  \textbf{\textminus0.46} &  \textbf{\textminus0.49} &  \textbf{\textminus0.51} &   \textbf{0.51} &   \textbf{0.47} &  \textbf{\textminus0.43} &  0.37 &   \textbf{0.39} & \textminus0.25 \\
    F & Thick--Thin       &   \textbf{0.45} &   \textbf{0.41} &   \textbf{0.44} &   \textbf{0.41} &  \textbf{\textminus0.43} &  \textbf{\textminus0.41} &   \textbf{0.35} & \textminus0.31 &  \textbf{\textminus0.33} &  0.07 \\
    G & Tense--Relaxed    &  \textbf{\textminus0.34} &  \textbf{\textminus0.29} &  \textbf{\textminus0.32} &  \textbf{\textminus0.37} &   \textbf{0.36} &   \textbf{0.37} &  \textbf{\textminus0.35} &   \textbf{0.34} &  0.19 & \textminus0.27 \\
    H & Dark--Bright      &   \textbf{0.53} &   \textbf{0.45} &   \textbf{0.52} &   \textbf{0.54} &  \textbf{\textminus0.54} &  \textbf{\textminus0.56} &   \textbf{0.50} &  \textbf{\textminus0.45} & \textminus0.34 & \textminus0.37 \\
    I & Cold--Warm        &   \textbf{0.36} &   \textbf{0.37} &   \textbf{0.36} &  \textbf{0.31} &  \textbf{\textminus0.33} &  \textbf{\textminus0.32} &   \textbf{0.22} & \textminus0.21 &  \textbf{\textminus0.28} & \textminus0.12 \\
        \bottomrule
      \end{tabular}%
    }
  \end{minipage}
\end{table*}
\subsection{Classical acoustic feature-based method}\label{sect:classical}
We first investigate how well classical acoustic features capture impression differences.
Acoustic features were extracted using openSMILE with the \texttt{eGeMAPSv02}~\cite{7160715} configuration, which has been widely used in paralinguistic and affective speech analysis tasks~\cite{app11188776}.
We used only features from voiced segments.

To represent the directionality between two utterances, we computed the feature difference 
$\boldsymbol{\phi}_\Delta = \boldsymbol{\phi}(x_b) - \boldsymbol{\phi}(x_a)$, which served as the input for regression models.
In a preliminary analysis, we examined correlations between $\boldsymbol{\phi}_\Delta$ and the annotated impression changes.
Several features showed moderate Pearson correlations with specific impressions.
Based on these correlations, we ranked all features and selected the eight most informative ones for each target impression.

\textbf{Regression models.}
The following regression models were used as a transparent and interpretable baseline for assessing how well classical acoustic features explain relative impression differences.
We trained Linear Regression (Linear), Ridge Regression (Ridge), Partial Least Squares (PLS2), Random Forest (RF), Gradient Boosting Decision Trees (GBDT), and Support Vector Regression (SVR).
Hyperparameters were determined experimentally, as summarized in Table~\ref{table:hyperparameters}.

\textbf{Neural model.}
To compare simple regression-based methods with a learnable nonlinear mapping, we trained a simple feed-forward neural network using a different input formulation.
The network received the concatenation of the selected features from both utterances, $[\boldsymbol{\phi}(x_a); \boldsymbol{\phi}(x_b)]$, instead of the difference vector $\boldsymbol{\phi}_\Delta$.
It consisted of three fully connected layers with ReLU activations and a final linear output layer of size $K=9$.
Training used mean squared error loss function and was optimized using the Adam optimizer~\cite{kingma-Adam} with a learning rate of 0.001 and a batch size of 8.
This setup allows the model to learn relative relations directly from both utterances.

\subsection{SSL-based method}\label{sect:ssl_based}

This experiment examined whether representations from SSL models can capture within-speaker relative impression changes while keeping the linguistic content constant.
In SSL-based modeling, frame-level representations are extracted from a pretrained encoder and aggregate into a fixed-size utterance-level vector~\cite{chen22g_interspeech,fujita2023zeroshot}.
Following this approach, each utterance $x_a$ and $x_b$ was encoded using a pretrained HuBERT~\cite{hsu2021hubert} model~(Fig.~\ref{fig:overview_propose}), pretrained on the ReazonSpeech corpus\footnote{\url{https://huggingface.co/rinna/japanese-hubert-base}}.
HuBERT produced 768-dimensional frame embeddings from \SI{16}{kHz} audio, which were summarized into 128-dimensional utterance embeddings via a weighted sum~\cite{chang22g_interspeech}, a bidirectional LSTM, and attention pooling~\cite{raffel2016feedforward}.
The embeddings $\boldsymbol{\psi}(x_a)$ and $\boldsymbol{\psi}(x_b)$ were concatenated and passed through a three-layer MLP to predict the nine-dimensional impression difference vector $\hat{\mathbf{r_{rel}}} \in \mathbb{R}^K$.
The model was trained using mean squared error loss and the AdamW optimizer~\cite{loshchilov2018decoupled} with a batch size of 8 and a learning rate of 0.002.

\subsection{MLLM-based method}\label{sect:mllm}
This experiment examined the zero-shot capability of MLLMs to reason about perceptual differences between two utterances.
Unlike the trainable classical and SSL-based approaches, this evaluation was exploratory rather than a strict benchmark, aiming to test whether recent audio-aware LLMs can perform pairwise impression analysis without additional training.

Each utterance pair was presented to an MLLM capable of directly processing audio inputs, with a detailed prompt instructing the model to assess perceptual differences along nine impression dimensions (e.g., Dark--Bright, Tense--Relaxed), as illustrated in Fig.~\ref{fig:overview_mllm}.
The model returned nine numerical scores and a brief textual rationale, which were parsed into a numerical impression-difference vector for quantitative analysis.
A prompt example is available on our demo page\footnote{\url{https://ntt-hilab-gensp.github.io/sp2026rie/}}.

We evaluated two models, ChatGPT (GPT-5, Thinking)~\cite{gpt5} and Gemini 2.5 Pro~\cite{comanici2025gemini25pushingfrontier}, both supporting direct audio input.
All inferences were performed through their respective web interfaces using default parameters.
Prompts were written in Japanese to match the dataset language, and evaluations were conducted in an in-context few-shot setting using a small number of scored example utterances as demonstrations.

\section{Results}
\begin{table}[tb]
 \centering
  \begin{minipage}{1.0\linewidth} 
    \caption{Pearson correlation coefficients among dimensions of impression difference vector. Values in bold show the optimal scores. }
    \setlength{\tabcolsep}{2pt}
    \label{table:peason}
    \centering
    \resizebox{\linewidth}{!}{%
    \begin{tabular}{@{}c@{)\ }l*{8}{S}@{}}
    \toprule
      \multicolumn{2}{c}{}&\multicolumn{7}{c}{Acoustic Features}&\multicolumn{1}{c}{SSL}\\
      \cmidrule(r){3-9} \cmidrule(l){10-10}
      \multicolumn{2}{c}{}&\multicolumn{1}{c}{Linear} & \multicolumn{1}{c}{Ridge}  &\multicolumn{1}{c}{PLS2}    &\multicolumn{1}{c}{RF}    &\multicolumn{1}{c}{GBDT}    &\multicolumn{1}{c}{SVR}   &\multicolumn{1}{c}{Neural}
      &\multicolumn{1}{c}{Neural}\\ 
      \midrule
      A& High--Low&0.54 & 0.54 &0.54 &0.54 & 0.51 & 0.49 &0.53& \textbf{0.60}\\ 
      B& Clear--Hoarse& 0.53 & 0.53 & 0.53 & 0.51 & 0.41 & 0.48 &0.58& \textbf{0.69}\\
      C& Calm--Restless& 0.32 & 0.32 &0.33 & 0.31 & 0.47 & 0.35 &0.55 & \textbf{0.65}\\
      D& Powerful--Weak& 0.42 & 0.42 & 0.42 & 0.43 & 0.41 & 0.40 &0.64& \textbf{0.72}\\
      E& Youthful--Elderly& 0.55 & 0.55 & 0.54 & 0.33 & 0.36 & 0.55 &0.58& \textbf{0.69}\\
      F& Thick--Thin& 0.45 & 0.47 & 0.46 & 0.44 & 0.41 & 0.41 &0.54& \textbf{0.59}\\ 
      G& Tense--Relaxed& 0.40 & 0.40 & 0.43 & 0.45 & 0.43 & 0.41 & 0.63&\textbf{0.66}\\ 
      H& Dark--Bright& 0.60 & 0.60 & 0.60 & 0.61 & 0.47 & 0.62 &0.72& \textbf{0.84}\\ 
      I& Cold--Warm & 0.38 & 0.38 & 0.38 & 0.38 & 0.43 & 0.31&0.49&  \textbf{0.69} \\
    \bottomrule
    \end{tabular}%
    }
  \end{minipage}
\end{table}

\begin{table}[tb]
 \centering
  \begin{minipage}{1.0\linewidth} 
    \caption{Concordance correlation coefficients among dimensions of impression difference vector. Values in bold show the best scores. }
    \setlength{\tabcolsep}{2pt}
    \label{table:ccc}
    \centering
    \resizebox{\linewidth}{!}{%
    \begin{tabular}{@{}c@{)\ }l*{8}{S}@{}}
    \toprule
      \multicolumn{2}{c}{}&\multicolumn{7}{c}{Acoustic Features}&\multicolumn{1}{c}{SSL}\\
      \cmidrule(r){3-9} \cmidrule(l){10-10}
      \multicolumn{2}{c}{}&\multicolumn{1}{c}{Linear} & \multicolumn{1}{c}{Ridge}    &\multicolumn{1}{c}{PLS2}    &\multicolumn{1}{c}{RF}    &\multicolumn{1}{c}{GBDT}    &\multicolumn{1}{c}{SVR}    &\multicolumn{1}{c}{Neural}&\multicolumn{1}{c}{Neural}\\ 
      \midrule
      A& High--Low&0.44 &0.44 & 0.45 & 0.47 & 0.43 &0.46 &0.41& \textbf{0.51}\\ 
      B& Clear--Hoarse& 0.44& 0.44 & 0.43 & 0.44 & 0.34 & 0.44 &0.50& \textbf{0.63}\\
      C& Calm--Restless& 0.20 & 0.20 & 0.20 & 0.24 & 0.39 & 0.30 & 0.44&\textbf{0.59}\\
      D& Powerful--Weak& 0.30 & 0.30 & 0.30 & 0.34 & 0.36 & 0.36 & 0.55& \textbf{0.67}\\
      E& Youthful--Elderly& 0.46 & 0.46 & 0.46 & 0.48 & 0.30 & 0.51 & 0.48&\textbf{0.64}\\
      F& Thick--Thin& 0.35 & 0.35 & 0.32 & 0.35 & 0.36 & 0.34 & 0.44&\textbf{0.50}\\ 
      G& Tense--Relaxed& 0.28 & 0.28 & 0.28 & 0.34 & 0.36 & 0.37 & 0.53&\textbf{0.60}\\ 
      H& Dark--Bright& 0.53 & 0.53 & 0.52 & 0.54 & 0.41 & 0.59 & 0.66&\textbf{0.82}\\ 
      I& Cold--Warm & 0.25 & 0.25 & 0.25 & 0.29 & 0.35 & 0.27 & 0.37&\textbf{0.64} \\
    \bottomrule
    \end{tabular}%
    }
  \end{minipage}
\end{table}

\begin{table}[tb]
 \centering
  \begin{minipage}{1.0\linewidth} 
    \caption{Pearson and concordance correlation coefficients resulting from MLLM methods, classical acoustic features, and SSL-based method. Values in bold show the optimal scores. }
    \setlength{\tabcolsep}{2pt}
    \label{table:results_mllm}
    \centering
    \resizebox{\linewidth}{!}{%
    \begin{tabular}{@{}c@{)\ }l*{4}{S}|*{4}{S}@{}}
    \toprule
    \multicolumn{2}{c}{}&\multicolumn{4}{c}{Pearson}&\multicolumn{4}{c}{CCC}\\
     \cmidrule(r){3-6} \cmidrule(l){7-10}
      \multicolumn{2}{c}{}&\multicolumn{1}{c}{Linear} & \multicolumn{1}{c}{SSL}    &\multicolumn{1}{c}{GPT}    &\multicolumn{1}{c}{Gemini}&\multicolumn{1}{c}{Linear} & \multicolumn{1}{c}{SSL}    &\multicolumn{1}{c}{GPT}    &\multicolumn{1}{c}{Gemini}\\ 
      \midrule
      A& High--Low&0.60&\textbf{0.66}&0.10&0.46&0.51&\textbf{0.63}&0.15&0.33\\ 
      B& Clear--Hoarse&0.59&\textbf{0.72}&0.19&0.41&0.51&\textbf{0.70}&0.19&0.30\\
      C& Calm--Restless&0.37&\textbf{0.73}&0.15&0.32&0.23&\textbf{0.67}&0.09&0.19 \\
      D& Powerful--Weak&0.37&\textbf{0.72}&\textminus0.04&0.35&0.31&\textbf{0.70}&\textminus0.03&0.22\\
      E& Youthful--Elderly&0.44&\textbf{0.64}&0.10&0.26&0.35&\textbf{0.58}&0.05&0.23\\
      F& Thick--Thin&0.48&\textbf{0.66}&0.15&0.21&0.37&\textbf{0.61}&0.12&0.14\\ 
      G& Tense--Relaxed&0.31&\textbf{0.64}&0.00&0.34&0.25&\textbf{0.62}&0.04&0.19\\ 
      H& Dark--Bright&0.69&\textbf{0.88}&0.19&0.56&0.62&\textbf{0.87}&0.16&0.45\\ 
      I& Cold--Warm &0.43&\textbf{0.75}&0.11&0.45&0.28&\textbf{0.69}&0.10&0.35\\
    \bottomrule
    \end{tabular}%
    }
  \end{minipage}
\end{table}

\subsection{Estimation with classical acoustic feature-based methods}
We evaluated both regression- and neural-based estimators using classical acoustic features extracted with openSMILE.
Performance was assessed using the Pearson correlation coefficient and the concordance correlation coefficient (CCC).

To select input features, we examined Pearson correlations between each openSMILE feature and the labeled impression differences.
The eight features showing the highest absolute correlations were selected, as summarized in Table~\ref{table:top_features}.
Among them, F0-related features, the first and second MFCCs coefficients, alpha ratio, the Hammarberg index, F1 bandwidth, and spectral flux appeared in the top eight for at least one of the nine impressions.

Pitch-related features showed relatively strong correlations with impression changes, consistent with previous findings in SER~\cite{ELAYADI2011572}.
Correlations computed at the 20th, 50th, and 80th pitch percentiles were highest at the 80th percentile, suggesting that high-pitch portions of speech strongly influence perceived impressions.
The first and second MFCC coefficients were also selected, whereas the third and fourth were not.
This finding implies that the global spectral envelope is  more perceptually relevant than fine-grained spectral details.
Measures associated with voice clarity, such as the Hammarberg index and alpha ratio, also showed moderate correlations.

Features not listed in Table~\ref{table:top_features} exhibited absolute correlation values below 0.3, indicating little relationship with impression differences.
These included loudness-related features, which have been reported as influential in previous work~\cite{frick1985communicating,laukkanen1997perception}. 
This discrepancy suggests that restricting the data to same-speaker and same-text pairs may have reduced their effect.
Variance-based statistics of the top features also showed weak correlations, while mean values were more informative.

\textbf{Regression models.}
As shown by the regression results in Tables~\ref{table:peason} and~\ref{table:ccc}, dimensions such as High--Low, Clear--Hoarse, Youthful--Elderly, and Dark--Bright achieved moderate performance.
Other dimensions showed correlations below about 0.4, indicating limited predictability.
The dimensions that were relatively well predicted corresponded to those strongly correlated with pitch-related descriptors, suggesting that pitch height plays a dominant role in these perceptual contrasts.
In contrast, impressions involving broader expressive qualities, such as Calm--Restless, Tense--Relaxed, Powerful--Weak, Thick--Thin, and Cold--Warm, were more difficult to estimate.
These impressions likely depend on multiple acoustic cues beyond pitch or other simple descriptors.

\textbf{Neural model.}
We also evaluated a simple feed-forward neural network that received the concatenation of the ten selected openSMILE features in Table~\ref{table:top_features} from both utterances as input.
Compared with regression-based models, the neural estimator achieved higher correlations overall across most impression dimensions.
Notably, dimensions such as Calm--Restless, Powerful--Weak, and Cold--Warm, which were poorly predicted by regression models, showed clear improvements with the neural approach.
Nevertheless, the average CCC values remained around 0.55 for most dimensions, indicating that estimation accuracy is still insufficient.
These results suggest that nonlinear mapping helps capture part of the perceptual relationships between utterances, but purely acoustic features alone may not fully explain subtle impression changes.

\subsection{Estimation with SSL-based methods}
We next evaluated the SSL-based model using the same metrics (Pearson correlation coefficient and CCC).
Results are shown in the SSL column of Tables~\ref{table:peason} and \ref{table:ccc}.
Across all impression dimensions, the SSL-based model outperformed the regression models using openSMILE features.

Notably, dimensions such as Calm--Restless, Powerful--Weak, and Cold--Warm, which had correlations below 0.4 with classical acoustic features, showed much higher correlations with SSL representations.
Several dimensions, including Powerful--Weak and Dark--Bright, achieved correlations exceeding 0.7.
These improvements indicate that SSL representations, which encode both high-level and temporal characteristics of speech, capture subtle impression changes more effectively within the same speaker.

The SSL-based model performed particularly well on impressions such as Calm–Restless and Powerful–Weak.  
These impressions are less influenced by static features like mean F0 and are more dependent on dynamic, contextual prosodic information, including speech rate, pause frequency and duration, and temporal evolution of the spectral envelope.
HuBERT, pretrained via masked unit prediction, inherently learns such temporal dependencies~\cite{fujita2023zeroshot,hsu2021hubert}.  
In contrast, the utterance-level statistics provided by openSMILE provide only coarse representations of temporal structure, limiting the ability of classical regression models to capture complex perceptual impressions.

These findings support previous results on the effectiveness of SSL representations for perceptual modeling~\cite{9893562}.
While classical acoustic features remain interpretable and useful, our results highlight their limitations in estimating complex, high-level perceptual attributes such as voice impressions.

\subsection{Estimation with MLLM-based methods}
Lastly, We evaluated two MLLMs, GPT, and Gemini, using one representative fold (81 utterances) and the same metrics.
As shown in Table~\ref{table:results_mllm}, the overall performance of the MLLMs was comparable to, or slightly lower than, that of the regression models based on classical acoustic features.
Gemini achieved moderate correlations for several dimensions, exceeding 0.4 for Dark--Bright, whereas GPT-5 showed almost none.

These results suggest that current MLLMs are not yet effective for comparative two-utterance tasks such as relative impression estimation as suggested in~\cite{wang2025speechllmasjudgesgeneralinterpretablespeech}.
Although the performance remains limited, improvements in prompting or few-shot adaptation could increase sensitivity to subtle acoustic contrasts.
While recent work has explored MLLMs as ``judges'' for speaking styles~\cite{wang2025speechllmasjudgesgeneralinterpretablespeech, chiang2025audioawarelargelanguagemodels}, our results indicate that  SSL-based approaches still perform more reliably even with limited labeled data.
Developing models that can reason about relative acoustic differences remains an important direction for future multimodal understanding.

\section{Conclusion}
We introduced relative voice impression estimation (RIE) and evaluated three approaches: classical acoustic features, self-supervised speech representations, and MLLMs. 
Methods using SSL representations consistently outperformed methods with classic acoustic features in both Pearson correlation and CCC, showing particularly large improvements on impressions that are less directly linked to specific acoustic features, such as Calm--Restless and Cold--Warm.
The results of the MLLMs suggest that current models are not yet reliable for fine-grained pairwise audio comparison.
This study had two main limitations: the experiments were conducted using speech data from a single female speaker, and MLLMs were used without fine-tuning.
Future work will explore few-shot training for MLLMs and extend RIE to multi-speaker and cross-gender settings.
These extensions will introduce new challenges, including speaker normalization and the disentanglement of speaker identity from expressive style.
\bibliographystyle{IEEEtran}

{\footnotesize \bibliography{mybib}}

\end{document}